\documentclass[11pt,oneside]{article}
\usepackage{academicons, lastpage, booktabs}
\usepackage[dvipsnames]{xcolor}
\usepackage[english]{babel}
\usepackage{amsfonts,amssymb,amsmath,amsthm}
\usepackage[absolute,overlay]{textpos}
\usepackage{float}
\usepackage{fancyhdr, graphicx}
\usepackage{enumerate}
\usepackage[margin=1in]{geometry}
\usepackage{tikz}
\usepackage{layout}
\usepackage{titlesec}
\usepackage{titling}
\usepackage{authblk}
\usepackage{xcolor, colortbl}
\usepackage{geometry}
\usepackage{siunitx} 
\usepackage{hyperref}
\usepackage{eso-pic}
\usepackage{marginnote}
\usepackage{doi}

\usepackage[mathlines]{lineno}
\usepackage{svg}
\usepackage{appendix}
\usepackage[square,numbers]{natbib}
\definecolor{coolblack}{rgb}{0.0, 0.18, 0.30}
\definecolor{carmine}{rgb}{0.59, 0.0, 0.09}
\definecolor{myblue}{rgb}{0.0, 0.48, 0.65}
\definecolor{cadetgrey}{rgb}{0.57, 0.64, 0.69}
\definecolor{lightgray}{gray}{0.9}

\hypersetup{
    colorlinks=true,
    linkcolor=coolblack,
   filecolor=magenta,      
    urlcolor=cyan,
    citecolor=coolblack,
    pdfhighlight=/P,
                 }
                                  
\usepackage{libertinus}
\usepackage[T1]{fontenc}
\usepackage{orcidlink} 

\usepackage{caption}
\usepackage{fancyhdr}

\titleformat*{\section}{\Large\bfseries\color{black}}
\titleformat*{\subsection}{\large\bfseries\color{black}}
\titleformat*{\subsubsection}{\bfseries\color{black}}

\makeatletter
\renewcommand\@makefntext[1]{\leftskip=0em\hskip0em\@makefnmark#1}
\makeatother
\newcommand\blfootnote[1]{%
  \begingroup
  \renewcommand\thefootnote{}\footnote{#1}%
  \addtocounter{footnote}{-1}%
  \endgroup
}

\providecommand{\keywords}[1]
{
  \small	
  \textbf{\textsf{Keywords---}} #1
}

\renewcommand{\tableautorefname}

\newcommand{\mycomment}[1]{} 
\newcommand*\mytitle{Diagnostics of a multicusp-assisted inductively-coupled radio-frequency plasma source for plasma immersion ion implantation}
\newcommand*\myauthor{Moreno \textit{et al.}} 
\newcommand*\mycorrmail{michael.bradley@usask.ca\\
lenaic.couedel@usask.ca, lenaic.couedel@cnrs.fr}
\newcommand*\myDOI{10.46298/ops.16754} 
\newcommand*\myarxiv{2510.18384} 
\newcommand*\myHAL{hal-05322136}
\newcommand*\myzenodo{-} 
\newcommand*\myvolume{2} 
\newcommand*\myyear{2026}
\newcommand*\mynumber{1 } 
\newcommand*\mydaterec{October 21, 2025} 
\newcommand*\mydaterevised{January 6, 2026} 
\newcommand*\mydateaccepted{January 14, 2026} 
\newcommand*\mydatepublished{January 28, 2026}

\newlength{\myshift}
\usepackage[firstpage=true]{background}
\backgroundsetup{contents={\includegraphics[width=1.\paperwidth]{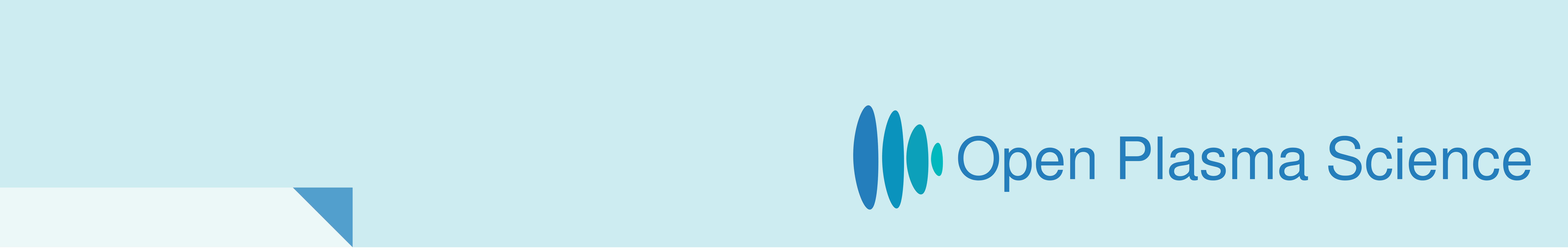}},scale=1,placement=top,opacity=1}

\pretitle{\vspace{-00pt}      \begin{flushleft}   \Large  \color{black} \selectfont  } 
     
\title{\textbf{\mytitle}}

\posttitle{\par \end{flushleft} } 

\preauthor{\vspace{-00pt} \begin{flushleft}  \color{black} \selectfont} 

\author[1]{Joel Moreno} 
\author[1]{Marilyn Jimenez}  
\author[1]{Daniel Okerstrom}  
\author[,1]{Michael P. Bradley\orcidlink{0000-0003-2412-4254}$^{*}$} 
\author[,1,2]{Lénaïc Couëdel\orcidlink{0000-0003-0749-9273}$^{*}$}

\affil[1]{\small Department of Physics and Engineering Physics, University of Saskatchewan, Saskatoon, SK S7N 5E2, Canada} 
\affil[2]{\small Aix-Marseille Université, CNRS, PIIM, UMR 7345, 13397 Marseille Cedex 20, France} 

\postauthor{\par\end{flushleft}} 

\date{}

\begin{document}

\maketitle
\thispagestyle{empty}
\blfootnote{$^*$ Corresponding authors: \textsf{\href{\mycorrmail}{\mycorrmail}}}
\blfootnote{Cite as: \myauthor, \mytitle, \textit{Open Plasma Science}  \myvolume , \mynumber (\myyear), doi: \myDOI}


\reversemarginpar

\marginnote{\begin{flushleft}
\sf \bf \footnotesize \color{coolblack} History
\end{flushleft}}[-\myshift]
\addtolength{\myshift}{-0.5cm} 
\marginnote{
\begin{flushleft} \footnotesize
Received \mydaterec
\end{flushleft}}[-\myshift]

\addtolength{\myshift}{-0.5cm} 
\marginnote{
\begin{flushleft} \footnotesize
Revised \mydaterevised
\end{flushleft}}[-\myshift]

\addtolength{\myshift}{-0.5cm} 
\marginnote{
\begin{flushleft} \footnotesize
Accepted \mydateaccepted
\end{flushleft}}[-\myshift]

\addtolength{\myshift}{-0.5cm} 
\marginnote{
\begin{flushleft} \footnotesize
Published \mydatepublished
\end{flushleft}}[-\myshift]

\addtolength{\myshift}{-1cm} 

\marginnote{\begin{flushleft}
\sf \bf \footnotesize \color{coolblack} Identifiers
\end{flushleft}}[-\myshift]
\addtolength{\myshift}{-0.5cm} 
\marginnote{
\begin{flushleft} \footnotesize
DOI \href{https://doi.org/\myDOI}{\myDOI} 
\end{flushleft}}[-\myshift]
\addtolength{\myshift}{-0.5cm} 
\marginnote{
\begin{flushleft} \footnotesize
HAL \href{https://hal.science/\myHAL}{\myHAL} 
\end{flushleft}}[-\myshift]
\addtolength{\myshift}{-0.5cm} 
\marginnote{
\begin{flushleft} \footnotesize
ArXiv \href{https://arxiv.org/\myarxiv}{\myarxiv} 
\end{flushleft}}[-\myshift]
\addtolength{\myshift}{-0.5cm} 
\marginnote{
\begin{flushleft} \footnotesize
Zenodo \href{https://https://zenodo.org/records/\myzenodo}{\myzenodo} 
\end{flushleft}}[-\myshift]

\addtolength{\myshift}{-1cm} 
\marginnote{\begin{flushleft}
\sf \bf \footnotesize \color{coolblack} Supplementary Material
\end{flushleft}}[-\myshift]
\addtolength{\myshift}{-0.5cm} 
\marginnote{
\begin{flushleft} \footnotesize
- 
\end{flushleft}}[-\myshift]

\addtolength{\myshift}{-1cm} 
\marginnote{\begin{flushleft}
		\sf \bf \footnotesize \color{coolblack} Licence
\end{flushleft}}[-\myshift]
\addtolength{\myshift}{-0.5cm}
\marginnote{
	\begin{flushleft} 
		\href{https://creativecommons.org/licenses/by/4.0/}{\includegraphics[width=0.55\marginparwidth]{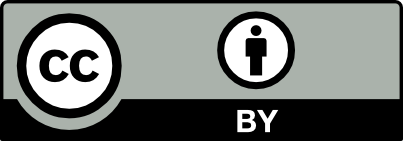}}
\end{flushleft}}[-\myshift]
\addtolength{\myshift}{-1.25cm} 
\marginnote{
	\begin{flushleft} \footnotesize
		\copyright  The Authors
\end{flushleft}}[-\myshift]

\addtolength{\myshift}{11.75cm} 
\marginnote{\begin{flushleft}
\sf \footnotesize \color{coolblack} \textsc{\textbf{Vol. \myvolume, No. \mynumber (\myyear)}}
\end{flushleft}}[-\myshift]


\addtolength{\myshift}{-1.5cm} 
  
\marginnote{\begin{flushleft}
\sf \normalsize \color{coolblack} \textsc{\textbf{Regular article}}
\end{flushleft}}[-\myshift]

\begin{flushleft}
\textbf{\textsf{Abstract}}
\vspace{5pt} 
\end{flushleft}

In this article, we present a detailed characterisation of a multicusp-assisted inductively coupled rf plasma source for plasma
immersion ion implantation (PIII). Using laser-induced fluorescence (LIF) and rf-compensated Langmuir probe diagnostics, we measured ion temperature $T_i$ and drift velocity $v_z$ in argon plasmas near an immersed electrode. The multicusp configuration enhances plasma density at low pressure, enabling stable operation down to 0.8\,mTorr. Time-averaged measurements show no detectable perturbation near the pulsed electrode, indicating full plasma recovery between high-voltage pulses. LIF-derived potential profiles match Riemann’s presheath theory, and ion velocity distributions reveal acceleration consistent with sheath dynamics. These results support the use of LIF for steady-state characterisation of the bulk and presheath regions in PIII systems.


\vspace{20pt} 


\begin{flushleft}

\keywords{Laser-indused fluorescence, inductively-coupled plasma} 

\end{flushleft}

\newpage

\newgeometry{ left=20mm, right=20mm,  bottom=3.5cm }

\pagestyle{fancy}
\setlength{\headheight}{15pt}
\fancyhead{} 
\fancyhead[L]{\footnotesize{\href{https://doi.org/\myDOI}{\myDOI}}} 
\fancyhead[R]{\footnotesize{\myauthor}}

\fancyfoot{} 
\fancyfoot[R]{\color{coolblack}{ \thepage \hspace{1pt} | \pageref{LastPage}}}
\fancyfoot[C]{\footnotesize{\color{black} Open Plasma Science \textbf{\myvolume}, No.\mynumber (\myyear)}}

\renewcommand{\headrulewidth}{0.4pt}
\renewcommand{\footrulewidth}{0.4pt}
\renewcommand{\footruleskip}{5pt}
\renewcommand\footrule{\hrule width\textwidth}
\renewcommand\headrule{\hrule width\textwidth}


\tableofcontents


\section{Introduction}
\label{sec:Intro}

Modern materials processing demands both precision and efficiency, which in turn require theoretical and computational models validated by empirical data. Among the various surface treatment technologies available, plasma-based processing has emerged as a leading approach due to its versatility and effectiveness across a wide range of applications. These include uniform thin-film deposition \cite{Jacob_2012, Mello_2011}, multi-tier nanostructure fabrication \cite{Bruchhaus_2017, Vempaire_2005}, high-aspect-ratio semiconductor etching \cite{Qian_1991, Yu_1994}, and improvements in surface durability and wear resistance \cite{Weng_2008, Park_2002}.

Historically, many techniques, such as conventional beamline ion implantation (CBII) \cite{Mantl_1992, Hutchings_1983}, chemical vapour deposition (CVD) \cite{Zhang_2013}, plasma polymerization \cite{bhatt2015}, plasma oxidation \cite{klages2023}, or physical vapour deposition (PVD) \cite{Geyao_2020}, have been widely used for surface modification. While these methods are still relevant, they present limitations. 
CBII, for instance, struggles with the uniform treatment of complex geometries and lacks the low-energy ion control needed for ultrashallow junctions in modern microelectronics \cite{Anders_2002,kim2017}. Vapour deposition techniques, meanwhile, often can suffer from isotropic etching and limited control over ion energy \cite{Lieberman,Abe2008,kouakou2007,Kumar2021}.

Plasma immersion ion implantation (PIII) offers a compelling alternative. It enables uniform treatment of irregular surfaces, independent control of ion fluence and impact energy, and a relatively simple system design \cite{chabert}. PIII has been successfully applied to tribological enhancement \cite{Ueda2005}, thin-film growth \cite{CatanioBortolan_2020}, etching \cite{Verdonck_2006, Donnelly_2013}, and the treatment of high-aspect-ratio silicon structures \cite{Jones1996}. A more recent application is the use of PIII to study ion bombardment damage to materials intended for fusion reactor plasma-facing components \cite{Yousaf2023}. In PIII, the target is immersed directly in the plasma, and ions are accelerated into the surface by the electric fields that naturally occur in the plasma sheath which surrounds the target. In most applications the target is pulsed with a succession of negative-polarity high-voltage (NPHV) pulses, although in some instances the target can be left floating, or be DC-biased. 
In the pulsed configuration, the ion fluence is determined by the bulk plasma charge density, while the ion impact energy is governed by the applied voltage \cite{Anders_2002}. Indeed, the electric field across the sheath between the bulk plasma and a target surface accelerates the ions for deposition and implantation in the PIII system. Applying NPHV pulses to the target modifies ion energy and fluence, enabling the greater control over implantation depth and ion flux. However, accurate modelling of this process is essential for optimising efficiency and yield.

The first model for HV sheath dynamics was proposed by Conrad \textit{et al.} \cite{Conrad1987} and later expanded by Lieberman \textit{et al.} \cite{Lieberman}. Lieberman’s model allows the estimation of ion fluence based on steady-state plasma parameters such as density and electron temperature. However, it assumes a stationary and spatially uniform plasma, and a sheath structure that remains consistent across pulses. These assumptions may not hold under realistic operating conditions.

In practice, the NPHV pulse can strongly perturb the plasma \cite{Moreno_2021}, even initiating breakdown in the absence of external sources, as seen in pulsed plasma thrusters and high-power impulse magnetron sputtering \cite{Gudmundsson2012, LING2020}. This challenges the assumption of steady-state plasma conditions and introduces variability in sheath properties such as potential drop, density, and spatial extent.

Several extensions to Lieberman’s model have been proposed, addressing finite pulse rise times \cite{Stewart1991}, sheath expansion across multiple pulses \cite{Emmert1992, wood_1993}, and collisional effects \cite{Wang1993}. Time-dependent sheath behaviour is often explored using Particle-in-Cell and Monte Carlo simulations \cite{wood_1993, keiter_1994, chung_1995}, which show that process efficiency is highly sensitive to pulse repetition rate \cite{tian_2000, daube_2001}. The repetition rate must be low enough to allow plasma recovery, yet compatible with thermal constraints imposed by target heating \cite{Anders}. A sufficiently dense plasma is less affected by the NPHV pulse, improving process reliability.

One effective route to increasing the background plasma density is the use of multicusp magnetic configurations \cite{Limpaecher1973,Jiang2020}. By arranging alternating‑polarity permanent magnets to form line or point cusps, electron confinement in the vessel is substantially enhanced due to the strong suppression of cross‑field wall losses (ions also benefitting when the magnetic field is sufficiently large). Such configurations are known to sustain large, uniform, and quiescent plasmas suitable for a wide range of processing and source applications \cite{Yamauchi1993,Kosonen2023,Bahari2026}. In the context of PIII, this approach is especially attractive: a denser and more stable plasma is noticeably less perturbed by the NHPV pulse, thereby improving overall process robustness.

This article presents the experimental characterisation of a multicusp-assisted, inductively coupled radio-frequency (rf) plasma source designed for PIII applications. The aim is to provide a stable, low-pressure plasma environment suitable for high-fidelity laser-induced fluorescence (LIF) diagnostics. Section~\ref{exp_setup} describes the experimental setup and the diagnostic tools. Section~\ref{Results} presents the results and the discussion. 

\section{Experimental setup}\label{exp_setup}

\subsection{Multicusp-assisted rf plasma source}\label{sec:source}

The plasma was generated using an inductively coupled radio-frequency (ICP) source operating at 13.56~MHz, designed by Plasmionique Inc.~\cite{bradley_2006} for the University of Saskatchewan Plasma Physics Laboratory. 
The system comprises a stainless steel vacuum chamber, rf power supply with matching network, a pumping system, and a high-voltage pulsing electrode for plasma immersion ion implantation (PIII) studies 
\cite{ayub_2019,Moreno_2021,Moreno_Thesis,Yousaf2023}.

The cylindrical chamber measures 29.8~cm in diameter and 46.0~cm in height, and includes eight access ports fitted with Conflat$^{\text{TM}}$ flanges for diagnostics such as Langmuir probes and optical viewports. To enhance plasma confinement, two rows of eight permanent ferrite magnets were mounted externally with alternating polarity, creating a multicusp magnetic field ranging from approximately 18~G at the chamber edge to $\lesssim$2~G at the centre.

A stainless steel electrode was inserted into the plasma for ion implantation experiments. With a surface area of 36.45~cm\textsuperscript{2}, it could be grounded, DC-biased, or pulsed using a custom-built Marx-bank modulator. For the experiments described, the electrode was directly exposed to the plasma to allow direct ion bombardment. It could be manually translated to adjust its depth within the plasma. The high-voltage pulses were applied at 250~Hz with durations of 40~$\mu$s, corresponding to a 1\% duty cycle, using a custom Marx-generator type high-voltage modulator \cite{Steenkamp_2007}. A pulse amplitude of  -1000~V was used. A laboratory power supply was used for measurements taken with the electrode biased to $-30$~V~DC.

Argon was used as the working gas, with flow rates between 0.5 and 1.0~sccm regulated by an MKS mass flow controller. 
The vacuum system consisted of a Pfeiffer HiPace 80 turbomolecular pump and a Leybold Turbovac 50, achieving a base pressure of 1~$\mu$Torr prior to plasma ignition. Operating pressures ranged from 0.7 to 2.0~mTorr, chosen to maintain high-confinement (H-mode) conditions. 

Plasma was ignited using a 600~W rf power supply connected to a water-cooled planar coil antenna. Impedance matching was handled by a Match Pro CPM-1000 network, which used a combination of fixed and variable capacitors to maintain a 50~$\Omega$ input impedance. Reflected power was kept below 5~W and the forward rf power was varied between 350 and 525~W to maximise fluorescence signal intensity.

\subsection{Langmuir probe}\label{sec:exp:LP}

Plasma parameters, such as the  electron and ion densities ($n_e$, $n_i$), electron temperature ($T_e$), and plasma potential ($V_{pl}$), were measured using a rf-compensated cylindrical Langmuir probe (LP). The probe design follows the rf-compensated model developed by Takashi \emph{et al.} for rf-driven plasmas \cite{takashi_2007}. The LP uses a tungsten tip, 0.15~mm in diameter and 6.37 mm in length, chosen to ensure sufficient current collection even in low-density regions of the plasma.

To minimise rf interference, the probe is equipped with four inductors tuned to the first two harmonics of the rf coil (13.56 MHz and 27.12 MHz), acting as band-stop filters. These reduce rf-induced fluctuations of the plasma potential, which can otherwise lead to overestimated electron temperatures \cite{chen_2003}. 
Additionally, a copper reference electrode is coupled to the probe tip via an 18 nF capacitor to further suppress rf noise and improve measurement accuracy \cite{chatterton_1990}.

\begin{figure}[h!]
	\centering
	\includegraphics[height=10cm]{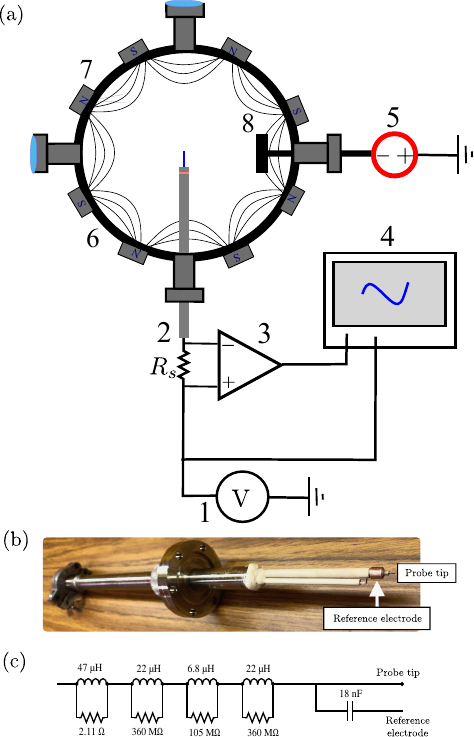}
	\caption{(a) Schematic of the experimental setup for Langmuir probe measurements: 1-- Langmuir probe controller adjustable DC power supply; 
		2-- Langmuir probe;  3-- Operational amplifier; 4-- Oscilloscope; 5-- High voltage pulser; 6-- Vacuum vessel; 7-- Multicusp belt of permanent magnets;
		8-- Stainless-steel electrode (substrate holder).
		(b) Picture of the Lanmuir probe.
		(c) Schematic of the probe’s internal circuitry, with four inductors connected in series to reduce the effect of the rf fluctuations on the measured signal. }
	\label{fig:LP_setup}
\end{figure}

The probe tip was placed along the central axis of the chamber about 15~cm below the rf antenna (see Fig.~\ref{fig:LP_setup}).
Plasma parameters were extracted from the probe IV curves using standard Langmuir analysis \cite{Merlino_2007,ayub_2019}, which assumes a Maxwellian electron energy distribution. This method was found to be accurate for ICPs in this pressure regime \cite{sukhanov_2004}, and compared favourably with other techniques such as OML \cite{Allen1992}, ABR \cite{Allen_1957}, and Druyvesteyn \cite{Godyak1993}, while being computationally efficient for processing large datasets. More details about the probe construction and characteristics can be found in Ref.~\cite{Moreno_Thesis}.

\subsection{Laser-induced fluorescence system}

\begin{figure*}[!htb]
	\centering
	\includegraphics[width = 0.99\textwidth]{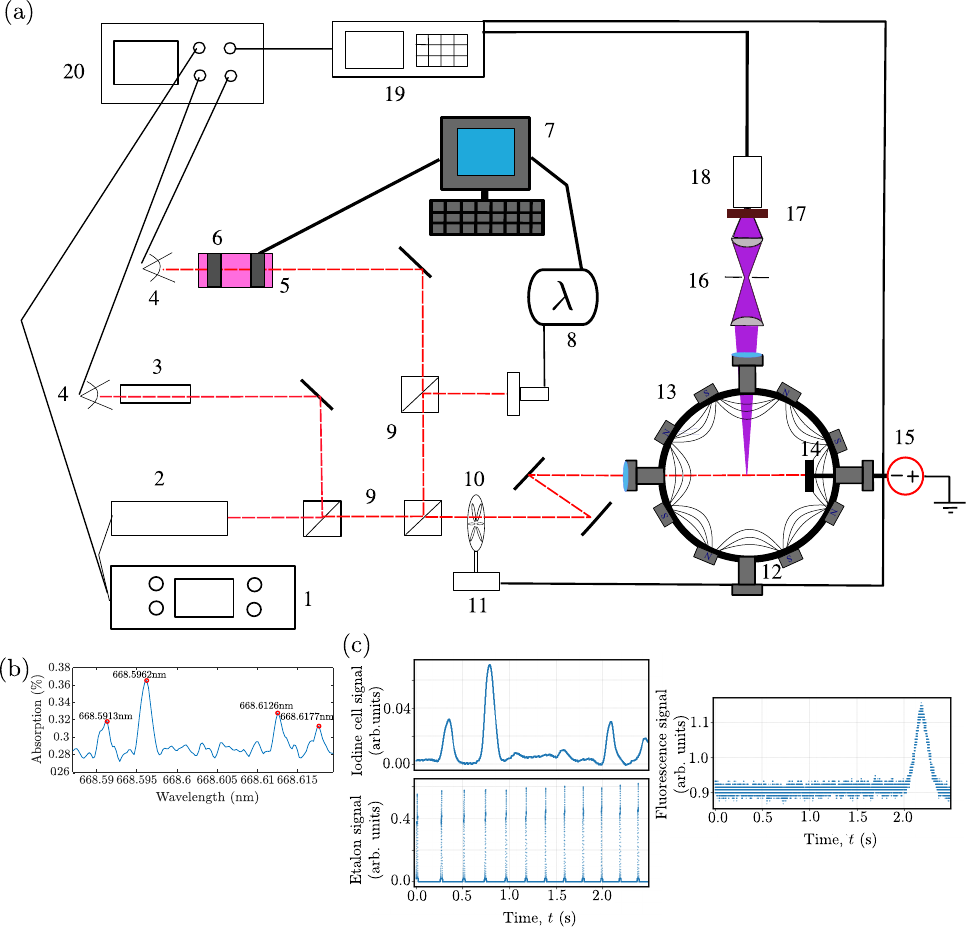}
	\caption{(a) Schematic of the experimental setup for LIF measurements. 1-- Laser controller; 2-- Laser head; 3-- Fabry-Perot etalon; 4-- Photodiode; 5-- Iodine cell; 6-- Iodine cell heater; 7-- Computer; 8-- Fibre optic head and wavemeter; 9-- Beamsplitter; 10-- Chopper blade; 11-- Chopper controller; 12-- Plasma chamber; 13-- Multicusp belt of permanent magnets; 14-- Pulsing HV electrode; 15-- Electrode NPHV pulsing system; 16-- LIF collection optics; 17-- Narrow bandpass optical filter; 18-- Photomultiplier tube; 19-- Lock-in amplifier; 20-- Oscilloscope. (b) Theoretical iodine absorption spectrum around the 668.6138~nm argon ion transition \cite{Salami_2005}. (c) Raw signal from the LIF setup. From top to bottom: Iodine cell absorption spectrum, Fabry-Perrot etalon signal, LIF signal.}
	\label{fig:LIF_setup}
\end{figure*}

Figure~\ref{fig:LIF_setup} shows the apparatus used for laser-induced fluorescence (LIF) measurements. This system consists of a tunable diode laser, calibration tools, and signal collection optics. The LIF system is designed to: i) calibrate the laser's relative frequency axis, and ii) detect and amplify the fluorescence signal.


\begin{figure}[h!]
	\centering
	\includegraphics[height=3cm]{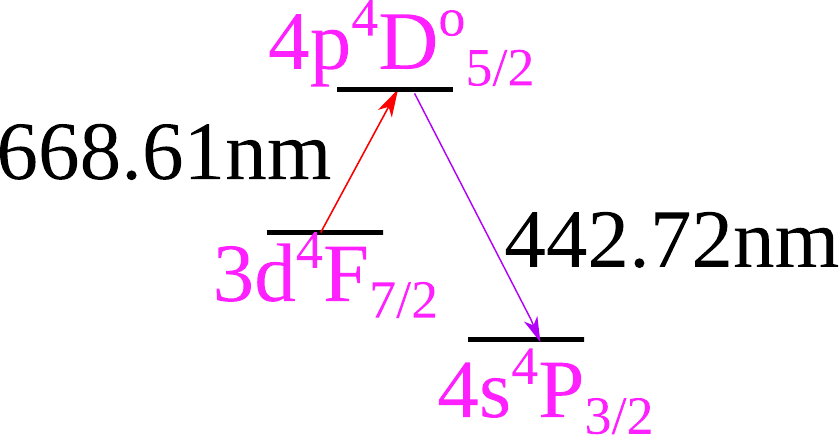}
	\caption{Partial Grotrian diagram for Ar-II transitions.}
	\label{fig:LIF_schemes}
\end{figure}

The light source used is a Toptica DL Pro diode laser, tunable in the range of 659–676\,nm, with a linewidth of 600\,kHz. The laser wavelength is controlled via adjustments to the grating angle, piezoelectric voltage, diode temperature, and current. A mode-hop-free scan over 20\,GHz ensures stable and continuous operation. Laser absorption occurs when the wavelength matches the transition energy of a specific excited ion state in its rest frame. The resulting emission intensity from the newly excited state is recorded as a function of the laser wavelength. Accurate wavelength-resolved intensity measurements enable measurement of the Ion Velocity Distribution Function (IVDF) via the Doppler-shift. The second moment of the IVDF provides a measure of the ion temperature $T_i$, while the peak position of the absorption profile yields the fluid velocity $v_i$. In our experiments, the laser was tuned to 668.6138\,nm, corresponding to the $\rm 3d^4 F_{7/2} \rightarrow 4p^4 D_{5/2}$ transition of Ar$^{+}$ (see Fig.~\ref{fig:LIF_schemes}) \cite{Keesee_2004} and the laser beam was directed perpendicularly to the surface of the stainless-steel electrode.

Calibration is performed using a wavemeter (Bristol 671B, 0.5 pm accuracy) and an iodine reference cell (I-cell). 
The I-cell is heated to $\sim80\ ^{\circ}$C to increase the iodine vapour density, enhancing absorption. Four iodine lines near 668.6138 nm (Fig.~\ref{fig:LIF_setup}(b) and (c)) assist in wavelength calibration \cite{Salami_2005}.

Relative frequency calibration is achieved using a home-made Fabry-Perot etalon, which acts as an air-spaced optical cavity. The etalon consists of two confocal mirrors with $\sim$99\% reflectivity at 680 nm and a radius of curvature of 44 mm. The mirrors are mounted on translation stages and separated by $4.2\pm0.1$~cm, giving a free spectral range $\Delta \nu_{FSR} = \frac{c}{4 \eta d}\sim1.78$~GHz, where $c$ is the speed of light in vacuum,  $d$ is the mirror separation and $\eta\simeq 1$ is the refractive index of air. The output intensity pattern is detected using a silicon photodiode (Thorlabs FD11A), housed in a metal case with a dedicated battery supply to minimise electrical noise (see Fig.~\ref{fig:LIF_setup}(c)). The etalon is enclosed in thermal and electrical shielding to ensure stability. The same photodiode setup is used for the iodine cell. Irregularities in the etalon signal indicate mode-hops, prompting further laser tuning.


Detecting fluorescence from optically-pumped plasma ions requires precise filtering and amplification due to the strong background plasma light and electrical noise from the rf generator. The signal is proportional to the density of argon ions in the metastable 3d $^4$F$_{7/2}$ state, which is populated via electron-impact excitation. The use of multicusp magnetic confinement using belts of alternating-pole permanent magnets allows us to reach relatively high ion density without at low pressure (which minimises quenching) (see Sec.~\ref{sec:source}). Previous studies showed that Zeeman splitting due to a magnetic field of 1 kG caused a broadening of 3.5 pm on the LIF signal \cite{Boivin_2003}. Assuming the splitting scales with the applied external magnetic field, this would result in a broadening of 0.007 pm in our set-up, which is a negligible affect \cite{telle_chem}.

The laser beam is modulated by an optical chopper (3.7~kHz) before entering the chamber through anti-reflective coated windows. To minimise reflection from the electrode, the beam is angled slightly off-normal. Fluorescent light is collected using plano-convex lenses and passed through an iris and a narrow bandpass filter centred on the wavelength of the emitted fluorescence photon (442~nm, FWHM 1.0 nm, 45\% transmission). The filtered light is detected by a Hamamatsu R3896 photomultiplier tube (PMT), which outputs to a lock-in amplifier (LIA) phase-locked to the chopper frequency. The LIA reduces noise and enhances signal sensitivity. All signals—including from the etalon, iodine cell, laser controller, and LIA—are recorded and averaged by a four-channel oscilloscope to produce the final raw signal [see Fig.~\ref{fig:LIF_setup}(c)]. 

LIF measurements were performed for pressure between 0.7~mTorr and 2~mTorr since pressures above 3~mTorr led to collisional quenching and the disappearance of the LIF signal, while pressures below 0.7~mTorr caused plasma instability and arcing during HV pulsing rendering the measurements unreliable.

\section{Results}\label{Results}

\subsection{Bulk plasma measurements: power and pressure dependence}\label{sec:bulk}

\begin{figure}[h!]
	\centering
	\includegraphics[width = 0.6\linewidth]{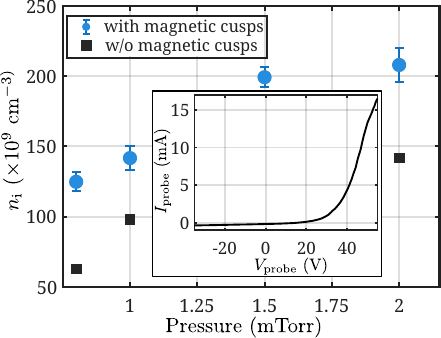}
	\caption{Ion density $n_i$ with (blue circles) and without (black square) magnetic confinement was 
		measured at various pressures at 500 W input power using a Langmuir probe positioned along the discharge axis. 
		For improved visual clarity, error bars for the measurements without magnetic cusps are omitted; however, they are comparable in magnitude to those 
		associated with the measurements taken in the presence of magnetic cusps. The inset shows a Langmuir probe characteristics at a pressure of 0.80~mTorr.}
	\label{fig:LP_MagvsNM}
\end{figure}

Langmuir probe measurements conducted at a constant forward rf power of 500~W but varying pressure indicate that the plasma ion density is significantly enhanced in the presence of magnetic cusps. At 0.8~mTorr, the plasma density is almost doubled --- from $\sim 6\times 10^{10}\ {\rm cm^{-3}}$ to $\sim 1.2\times 10^{11}\ {\rm cm^{-3}}$ (see Fig.~\ref{fig:LP_MagvsNM}) --- demonstrating the improved confinement of the plasma resulting from the introduction of magnetic cusps. This result was expected from the literature \cite{Limpaecher1973,Jiang2020,Yamauchi1993,Kosonen2023,Bahari2026}. It enabled stable plasma operation of our device at low pressures, down to 0.8~mTorr. However, the enhanced confinement due to magnetic cusps diminishes at higher pressures as a result of increased collisionality \cite{Jiang2020}.

\begin{figure*}[h!]
	\centering
	\includegraphics[width=0.75\textwidth]{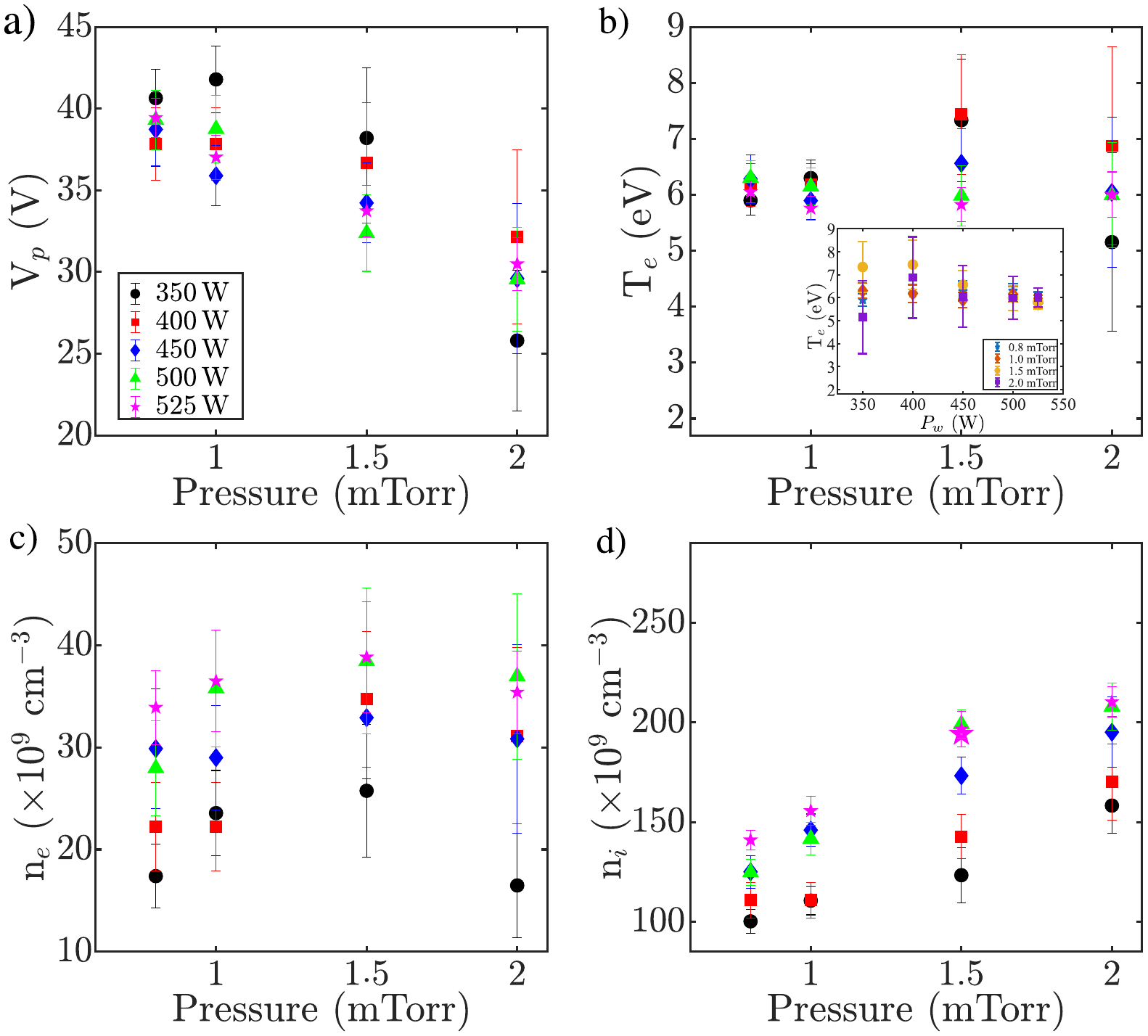}
	\caption{Langmuir probe measurements as a function of rf power and pressure: 
		(a) plasma potential \(V_{\text{pl}}\), (b) electron temperature \(T_e\), (c) electron density \(n_e\), and (d) ion density \(n_i\). The electron density \(n_e\) is approximately five times lower than the ion density \(n_i\), likely due to probe-induced perturbations that drain the plasma and alter its equilibrium state \cite{chen_2003}. For clarity, the inset in (b) also shows \(T_e\) as a function of the rf power $P_w$. }
	\label{fig:LP_results_bulk}
\end{figure*}

Then, the Langmuir probe measurements focused on characterising the bulk plasma (i.e., at the center of the vacuum chamber) under varying forward rf power (350--525~W) and neutral argon pressure (0.8--2.0~mTorr). These conditions were chosen to allow complementary LIF studies presented later in the section (in this pressure range, we avoid collisional quenching at high pressures (the LIF signal became very weak above 2.0~mTorr) and unstable H-mode of the ICP discharge at low pressures (below 0.8~mTorr)). Rf-compensated Langmuir probe measurements gave us $n_i$, $n_e$, $T_e$, and $V_{pl}$ (Figure~\ref{fig:LP_results_bulk}) while LIF measurements focused on $T_i$ and the ion drift velocity $v_z$. 

Langmuir probe measurements must be interpreted carefully. While electron saturation current $I_{e,sat}$ can indicate $n_e$, this is reliable only at low densities and pressures, where the electron mean free path is long \cite{chen_2003}. At higher densities, the probe can drain the plasma, altering its equilibrium. 
Additionally, due to experimental noise and probe contamination, the identification of $I_{e,sat}$ can be inaccurate \cite{Jin2023}. Therefore, ion current collection is often preferred for estimating density (here we used Orbital Motion Theory to extract plasma density from the ion current \cite{chen_2003}), especially given plasma quasineutrality \cite{chen_2003}. Additionally, due to the inherent noise of our data acquisition system, the error bars on the Langmuir probe results were quite large and the obtained values thus served as order-of-magnitude estimates only. The observed trends are (see Fig.~\ref{fig:LP_results_bulk}):

\begin{itemize}
	\item $V_{pl}$ decreases with pressure and is almost independant of the forward rf power.
	\item $n_i$ (and $n_e$) increase with power and pressure (\(n_e\) is approximately five times lower than \(n_i\) with much larger errorbars, likely due to noise in the IV characteristics and probe-induced perturbations that drain the plasma and alter its equilibrium state \cite{chen_2003}).
	\item Measured electron temperature ($T_e$) ranged from 5 to 7.5~eV and was found to remain relatively constant. 
\end{itemize}

Our findings are generally consistent with results obtained from Langmuir probe diagnostics reported in the literature~\cite{Lieberman,Godyak_2002}. Prior studies suggest that the electron temperature (\(T_e\)) is primarily governed by pressure, increasing as pressure decreases, while exhibiting only a weak dependence on rf power within certain operational regimes~\cite{Hori1996,Hopwood1993}. In our measurements, however, this trend is not clearly discernible due to the large error bars. In contrast, electron density (\(n_e\)) was found to be strongly influenced by both pressure and power, increasing with higher rf power and higher pressure~\cite{Hori1996}. Across multiple experiments, \(T_e\) typically ranged between 2--7~eV~\cite{Menard1996,Hopwood1993}. At low rf power, \(T_e\) could be smaller for the highest pressure (errorbars are too large to conclude), whereas at high rf power, there are no observable differences between pressure. The interplay between neutral gas heating and stepwise ionisation, which modifies the electron energy distribution, is known to play a role on the pressure/power dependence of the electron temperature~\cite{Lee2017a}.
Neutral gas dynamics also play a significant role: neutral density is often depleted in the plasma center due to heating from collisions with charged particles~\cite{Hori1996}. Furthermore, the plasma heating mechanism transitions from stochastic heating at low pressure to collisional heating at higher pressures. This shift affects how electrons gain energy, particularly in relation to nonlocal electron kinetics and ambipolar potential barriers~\cite{Karar2021}.

\begin{figure}[h!]
	\centering
	\includegraphics[width=0.5\linewidth]{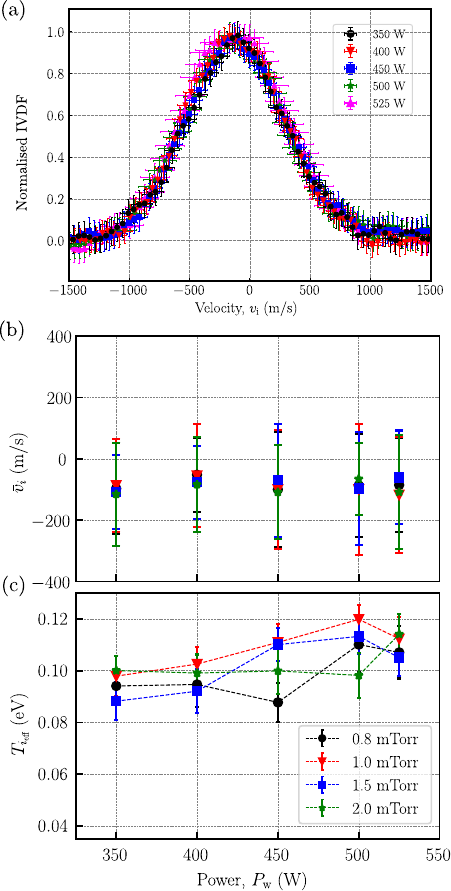}
	\caption{(a) Normalised IVDFs at 0.8~mTorr for various rf powers. 
		(b) Ion drift velocity as a function of power for different argon pressure. 
		(c) Ion temperature as a function of power for different argon pressure.}
	\label{fig:LIF_Bulk}
\end{figure}

The normalised Ion Velocity Distribution Functions (IVDFs) presented in Figure~\ref{fig:LIF_Bulk}(a) exhibit a discernible broadening as the applied power increases, indicative of a higher ion temperature ($T_i$). This interpretation is corroborated by Figure~\ref{fig:LIF_Bulk}(c), which demonstrates a slight increase in $T_i$ from  $\sim0.09~\mathrm{eV}$ to $\sim0.11~\mathrm{eV}$ over the investigated power range. The dependence of $T_i$ on pressure, also shown in Figure~\ref{fig:LIF_Bulk}(c), is unclear and no trend can be directly seen. These findings are broadly consistent with previous investigations, which reported ion temperatures in the range of $0.04$--$0.08~\mathrm{eV}$ and observed similar trends of increasing $T_i$ with power\cite{Hebner_1996,Sadeghi_1997,Jun_2006}. Jun \textit{et al.}~\cite{Jun_2006} reported that for pressures 
exceeding approximately $1.5~\mathrm{mTorr}$, $T_i$ begins to decrease but our current measurements are unable to confirm this finding. Therefore, the behaviour of $T_i$ at higher pressures remains an open question and merits additional study.

The directed (drift) ion velocity remains slightly negative, approximately $-75~\mathrm{m/s}$, across all investigated powers and pressures [Figure~\ref{fig:LIF_Bulk}(b)]. In an idealised symmetric 
discharge, the drift velocity at the plasma centre would be expected to be exactly zero. The observed deviation is likely attributable to asymmetries introduced by diagnostic access ports and other structural features of the vacuum vessel, as well as a potential lateral displacement of the LIF interrogation volume from the discharge axis.

\subsection{Influence of proximity to electrode (sheath profile)}

This section characterises the ion dynamics in the vicinity of an immersed electrode in our multicusp-assisted inductively coupled argon plasma, with three aims: (i) determine whether time-averaged LIF can reveal signatures of the high-voltage (HV) implantation sheath under negative-polarity pulses; (ii) map the directed ion velocity $v_z$ and infer the steady-state presheath/sheath potential profile for grounded and DC-biased conditions; and
(iii) examine the evolution of the ion temperature $T_i$ as the measurement volume approaches the electrode.

Experiments were performed at $P_{\rm RF}=500~\mathrm{W}$ and $p=0.8~\mathrm{mTorr}$. The electrode was set to: (a) ground, (b) $-30~\mathrm{V}$ DC, and (c) pulsed at $-1~\mathrm{kV}$ with $40~\mu\mathrm{s}$ duration at $250~\mathrm{Hz}$ (1\% duty cycle). Spatial scans were executed from $z=0.1$ to $2.0~\mathrm{cm}$ in front of the electrode (finer steps of $0.1~\mathrm{cm}$ within $0.1$–$0.5~\mathrm{cm}$).

\begin{figure}[h!]
	\centering
	\includegraphics[width=0.8\linewidth]{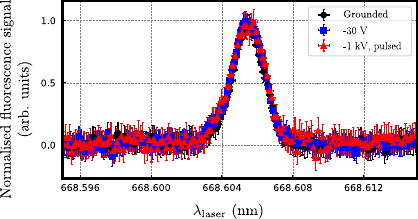}
	\caption{Normalised fluorescence profiles as a function of laser wavelength $\lambda_{\rm laser}$ measured at a distance of $1~\mathrm{mm}$ from the electrode surface. Three conditions are shown: grounded electrode (black), DC-biased electrode at $-30~\mathrm{V}$ (blue), and electrode subject to $-1~\mathrm{kV}$ pulsed bias (red). For clarity, the spectral resolution in $\lambda_{\rm laser}$ has been reduced in the plot.}
	\label{fig:LIF1mm}
\end{figure}

The time-averaged fluorescence profiles at 1~mm of the electrode surface for different bias are shown in  Fig.~\ref{fig:LIF1mm}. No detectable deformation under HV pulsing nor the DC bias can be seen. Consistently, $v_z(z)$ is essentially indistinguishable between grounded, $-30~\mathrm{V}$ DC, and pulsed case. Within the averaging limits, this supports the common approximation that bulk parameters between pulses are only weakly perturbed.

\begin{figure}[h!]
	\centering
	\includegraphics[width=0.7\linewidth]{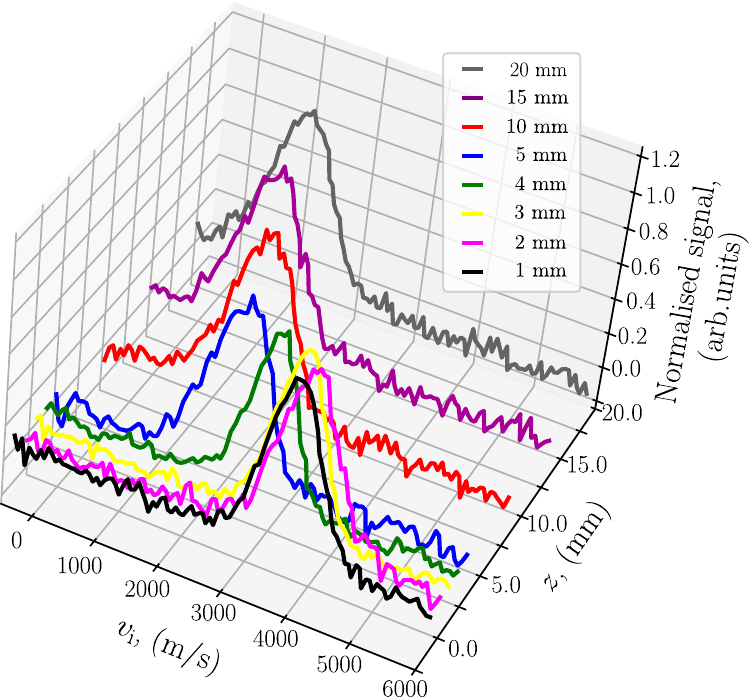}
	\caption{Normalised ion velocity distribution functions (IVDFs) measured at various distances $z$ from a 
		grounded electrode. The profiles illustrate the evolution of the IVDF shape and 
		peak position as the electrode is approached. For clarity, the errorbars ($\sim 5$\%) have been removed.}
	\label{fig:IVDFs}
\end{figure}

Figure~\ref{fig:IVDFs} presents the normalised experimental IVDFs at different distances $z$ from the grounded electrode. As $z$ decreases, the peak of the distribution shifts towards higher velocities, indicating ion acceleration towards the electrode in the presheath region. The IVDF width appears to decrease progressively down 
to approximately $z = 4~\mathrm{mm}$; beyond this point, the trend becomes less clear. The increase in ion velocity observed close to the target is attributed to ion acceleration by the electric field in the presheath region. The measurements concern metastable argon ions (Ar$^+$*, $3d^4 F_{7/2}$), which represent less than 1\% of the total ion population \cite{Sadeghi_1997}. Due to their short mean free paths, these ions are generated locally via electron impact, rather than being transported from outside the presheath. The absence of a zero-velocity component in the IVDFs confirms that excitation occurs after acceleration. As a result, the measured velocity distribution of metastable ions should accurately reflect that of ground-state argon ions \cite{Sadeghi_1997}. 

\begin{figure}[h!]
	\centering
	\includegraphics[width=0.5\linewidth]{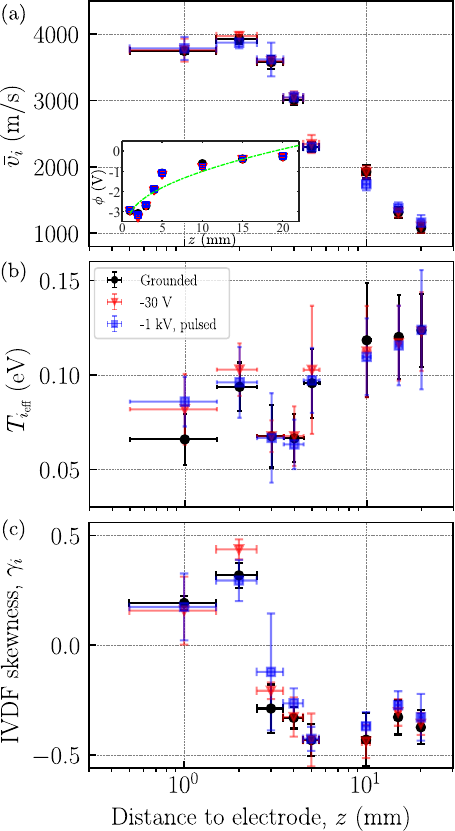}
	\caption{Spatial evolution of IVDF parameters as a function of distance 
		$z$ from the electrode: (a) mean ion velocity $\bar{v}_i$, 
		(b) effective ion temperature $T_{i,\mathrm{eff}}$, and 
		(c) skewness $\gamma_i$. Results are shown for grounded, 
		DC-biased, and pulsed-bias conditions. The inset in (a) shows the 
		electric potential profile assuming collision-less ions and the dashed green line is the fit using Eq.\ref{Eq:Riemann}.}
	\label{fig:paramIVDF}
\end{figure}

Equivalent measurements were performed for an electrode biased at $-30~\mathrm{V}$ and for an electrode subject to $-1~\mathrm{kV}$ pulsed bias (red curves). Each IVDF was fitted with a skewed Gaussian function to quantify asymmetry. From these fits, the mean ion velocity $\bar{v}_i$, the effective ion temperature $T_{i,\mathrm{eff}}$, and the skewness $\gamma_i$ were extracted as functions of $z$ (Fig.~\ref{fig:paramIVDF}). The profiles for all electrode biases are nearly identical, reinforcing the previously-noted assumption that bulk plasma parameters remain largely unperturbed between short-duty-cycle pulses.

The mean ion speed increases monotonically as $z$ decreases, approaching the theoretical Bohm velocity $u_B$. 
Based on Langmuir probe measurements at $p = 0.8~\mathrm{mTorr}$ and $P_{\rm RF} = 500~\mathrm{W}$ (Sec.~\ref{sec:bulk}), the Debye length is estimated as $\lambda_D \approx 40$--$80~\mu\mathrm{m}$, the Bohm velocity as $u_B \approx 3$--$4~\mathrm{km\,s^{-1}}$, and the steady-state sheath thickness as $S_{\rm is} \approx 0.3$--$0.6~\mathrm{mm}$~\cite{Moisan}. LIF measurements indicate that ions reach $u_B$ at $z \simeq 0.1$--$0.2~\mathrm{cm}$, which is slightly larger than the estimated sheath length but remains consistent within the large theoretical uncertainties and the modified conditions introduced by the presence of the electrode.

The fits of the IVDFs indicate that $T_i$ decreases from $z=20$ to $4~\mathrm{mm}$ and then increases from $d=4$ to $2~\mathrm{mm}$ (Fig.~\ref{fig:paramIVDF}(b)) and seems to decrease again at $z=1$~mm. This non-monotonic trend contrasts with several earlier reports of monotonically increasing $T_i$ towards a boundary in other devices/conditions \cite{Bachet_1995,Sadeghi_1991,Oksuz_2001,Oksuz_2005}. However, the neutral gas temperature profile in the discharge is unknown and it is reported in the literature that the neutral temperature has a maximum in the centre of the discharge and decreases towards the edge \cite{Hebner_1996}. Ion-neutral collision might therefore explain why the ion temperature decreases towards the electrode.

Figure~\ref{fig:paramIVDF}(c) presents the evolution of the skewness of the IVDF as a function of the distance $z$ from the electrode. Between $z = 20$~mm and $z = 5$~mm, the skewness is negative and decreases from approximately $0.35$ to $0.45$, indicating a slight deviation from a Maxwellian distribution. A negative skewness implies an asymmetry in the IVDF, with a higher population of slow ions compared to fast ions. This feature may be attributed to low-velocity ion populations generated by ionisation or charge-exchange processes~\cite{Claire_2006,Coulette_2015,Yip_2015}. As the distance to the electrode decreases further, the skewness increases, reaching zero at $z = 3$~mm (Maxwellian distribution) and becomes slightly positive at shorter distances, where the fast-ion side dominates. It is important to note that the spatial resolution ($\Delta z \sim 0.5$~mm) can introduce artificial broadening of the IVDF due to averaging over the detection volume. Near the target, where ion velocity increases rapidly, this effect may distort the measured IVDF towards higher velocities, leading to an overestimation of both the drift velocity and the ion temperature~\cite{Grift1997}.

\section{Discussion}

At a pressure of 0.8\,mTorr, the neutral density is approximately $n_n \approx 3 \times 10^{19}$\,m$^{-3}$. 
Assuming an ion-neutral collision cross section of $\sigma \sim 10^{-18}$\,m$^{2}$~\cite{Lieberman}, 
the corresponding ion mean free path is $\lambda_{i,\mathrm{MFP}} \sim 30$\,mm~\cite{Lieberman}.
This suggests that most LIF measurements are performed well within the spatial extent of the presheath. Using order-of-magnitude inputs from probe diagnostics at $0.8$\,mTorr and $500$\,W (Section~\ref{sec:exp:LP}), the following plasma parameters are also estimated: Debye length $\lambda_D \approx 50\,\mu$m, ion sound speed $u_B \approx 3.9$\,km\,s$^{-1}$, and ion sheath thickness $S_{\mathrm{is}} \sim 0.3$\,mm. Using energy conservation, the local potential can be inferred from the axial ion velocity via $\phi_{\mathrm{ps}}(z) = \tfrac{M_i \bar{v}_i^2}{2e}$. The resulting LIF-derived potential profile can be fitted reasonably well with Riemann’s 
analytical model for a weakly collisional presheath \cite{Riemann_1991,Lee2022}:

\begin{equation}\label{Eq:Riemann}
	\phi(z) = T_e \sqrt{\frac{(z-z_0)}{\lambda_{i,\mathrm{MFP}}}} + \phi_0,
\end{equation}

where $z_0$ is the position of the sheath--presheath boudary. The fit (inset of Fig.~\ref{fig:paramIVDF}(a)) yields an effective electron temperature of $T_e \approx 2.8$\,eV (within a factor of $\sim 2.2$ of probe (consistent with typical probe uncertainties) and close global model estimates ($\sim$2.5~eV using the model described in Ref.~\citep{Lieberman})), an ion neutral mean free path $\lambda_{i,\mathrm{MFP}}\sim 12$~mm (consistent with our previous estimate), and a sheath presheath boundary at $\sim 0.8$~mm (in reasonable agreement with our estimate). The measured ion drift velocity $v_z(z)$ reaching the Bohm speed $u_B$ and the closeness of the LIF-derived potential profile $\phi(z)$ with Riemann’s model confirm the successful access to the presheath region. 

The non-monotonic behaviour of $T_i(z)$ near the electrode likely results from a combination of factors:
measurement noise, deviations from Maxwellian IVDFs, metastable-state kinetics, and spatial variations in
neutral gas temperature. The observed IVDF skewness suggests the presence of low-velocity ion populations
due to ionisation and charge-exchange processes \cite{Bachet_1995,Sadeghi_1991,Oksuz_2001,Oksuz_2005}.

Comparisons between grounded, DC-biased, and NPHV-pulsed conditions show that steady-state plasma parameters remain largely unaffected by the NPHV pulses, consistent with the short duty cycle used. In the context of PIII, it tends to confirm that the use of multicusp magnetic configurations allowing a denser and more stable plasma is less perturbed by the NHPV pulse and is interesting for the process robustness. Nevertheless, for NPHV pulses of $V_s = -1000$\,V, the estimated sheath thickness ($\sim 2$\,mm) lies within the LIF collection volume. However, the corresponding Doppler shift ($\sim 100$\,GHz) exceeds our laser’s tuning range, limiting the direct observation of NHPV sheath dynamics. Time-resolved LIF with extended frequency access is needed to capture transient effects during NPHV pulses.

\section{Conclusion}

In this study, laser-induced fluorescence was used to investigate ion dynamics in the bulk and in the presheath regions of a multicusp-assisted ICP discharge. The system successfully measured $T_i$ and $v_z$ in both bulk plasma and near an immersed electrode.

The key findings are:
\begin{itemize}
	\item Ion temperature $T_i$ increases with rf power in the bulk plasma, consistent with prior studies. No dependance of $T_i$ on pressure could be clearly identified.
	\item Ion drift velocity $v_z$ remains near zero in the bulk but increases near the electrode, reaching $u_B$ at $\sim 1$\,mm from the electrode surface ($\gtrsim 20 \lambda_{\rm D}$).
	\item Potential profiles derived from LIF match relatively well the Riemann’s weakly collisional presheath model.
	\item No time-averaged distortion was observed under NPHV pulses, confirming plasma recovery between pulses and the suitability of a multicusp-assisted ICP discharge for reliable PIII.
\end{itemize}

Future work should focus on improving signal-to-noise ratio and improving the  diagnostics capabilities to improve the accuracy of the measurements and  enable time-resolved LIF and capture the NPHV sheath dynamics. Additional parameter scans and collisional–radiative modelling will help clarify ion temperature behaviour and metastable kinetics as a function of rf power and pressure. Complementary diagnostics and simulations will support benchmarking of sheath-edge localisation and potential profiles.


\section*{Acknowledgments}
\addcontentsline{toc}{section}{Acknowledgments}

The authors acknowledge the support of the Natural Sciences and Engineering Research Council of Canada (NSERC), Grants  No. RGPIN-2019-04333 and No. RGPIN-2025-05757.


\bibliographystyle{aomalpha}
\bibliography{LIF_Ref}
\addcontentsline{toc}{section}{References}

\end{document}